# Selective Harmonic Elimination in a Cascaded H-Bridge Multilevel Inverter fed by High-Frequency Isolated DC-DC Converter

Amirhossein Pourdadashnia
*Faculty of Electrical and Computer Engineering*
*Urmia university*
*Urmia, Iran*
st_a.pourdadashnia@urmia.ac.ir

Milad Sadoughi
*Faculty of Electrical and Computer Engineering*
*Urmia university*
*Urmia, Iran*
st_m.sadughi@urmia.ac.ir

Mohammad Farhadi-Kangarlu
*Faculty of Electrical and Computer Engineering*
*Urmia university*
*Urmia, Iran*
m.farhadi@urmia.ac.ir

Behrouz Tousi
*Faculty of Electrical and Computer Engineering*
*Urmia university*
*Urmia, Iran*
b.tousi@urmia.ac.ir

*Abstract*—In recent years, the use of multilevel inverter has become popular due to its many advantages. Due to the popularity of multilevel inverters in industries and in applications that require a wide range of voltages, there have been many challenges to achieve high-quality voltage. Many pieces of research have been done to solve the problem of annoying harmonics in multilevel inverters. In this study, pulse width modulation (PWM) has been proposed as a switching method. Using the Selective Harmonic Elimination (SHE) technique, the inverter switching can be carried out in low frequency, and also, this method can reduce the annoying harmonics significantly. However, in the lower modulation index, eliminating such harmonics is challenging, resulting in a considerable increase in output voltage distortion. This study suggests a way to solve the problem. In this research, eliminating selected harmonics in a multilevel inverter with variable DC links is proposed. The DC-link variable method in this study is to use a high-frequency isolated DC-DC converter. The proposed method is verified on a 5-level Cascaded H-Bridge (CHB) inverter using the SHE-PWM method solved by particle swarm optimization (PSO) algorithm.

## I. INTRODUCTION

Multilevel inverters (MLIs) have been used in high-voltage, high-power applications in various usage due to their capability to generate high-quality output waveforms with lower switching frequencies [1-3]. The MLIs have attracted much attention in medium and high voltage applications than two-level inverters under low switching losses, high effectiveness, and higher electromagnetic compatibility. Among several topologies of the MLIs, the cascaded H-bridge (CHB) inverter is considered a desirable option for providing better quality power [4-5].

The CHB inverter is superior to other topologies due to its modularity and ease of control. Several conventional methods for pulse width modulation (PWM), such as sinusoidal PWM carrier (SPWN) and space vector modulation (SVM), have been suggested to realize the output voltage control function and simultaneously reduce unwanted harmonics [6-7]. Low-frequency modulation techniques, such as selective harmonic elimination (SHE), are used in basic frequency operations to limit losses and increase converter efficiency. The SHE-PWM technique can produce high-quality output voltage with the advantage of lower switching frequency compared to other modulation methods [8]. The MLIs typically operate at low switching frequencies, allowing SHE-PWM to gain direct control of low-order harmonics more efficiently. Using the SHE-PWM technique aims to eliminate or minimize low-order harmonics [9]. To eliminate the low order harmonics from the output voltage waveform of a CHB MLI, at first, the switching angles are calculated by solving the SHE-PWM equations. The challenge of solving equations to find the switching angles in this research is solved by the robust particle swarm optimization (PSO) algorithm [10-14]. There are several optimization algorithms such as PSO algorithm, teaching learning-based optimization (TLBO) algorithm, genetic algorithm (GA), etc. can be used for efficiently solving the aforementioned equations [15-16]. However, the Newton-Raphson and GA methods require an initial guess, which must be very close to the exact solution. It is challenging to make a good guess in most cases. This is because the search space for the SHE-PWM problem is unknown. Therefore, PSO is proposed as a powerful algorithm in this research [17-18]. However, reaching the voltage with acceptable quality is a challenge, and despite using the SHE-PWM technique, there is still the problem of harmonics in low modulation index [19-21]. In the proposed method, at low output voltage amplitudes, the low-order harmonics are eliminated or attenuated, and as a result, the output voltage waveform, as well as THD amount, are significantly improved. In this method, a wide range of high-quality inverter output voltages is obtained. One of the advantages of the proposed method is reducing voltage stress on the switches and a considerable reduction in the size of the filter.

To improve the quality of the output voltage and the THD amount, a high-frequency isolated DC-DC is introduced to adjust the dc-link voltage. Also, the isolated high-frequency DC-DC converter utilizes high frequency (HF) switching to minimize the converter's size and weight, which in turn increases the power density of the converter. In this paper, using the PSO algorithm, the SHE-PWM equations are solved, and the proposed method for a 5-level CHB MLI is Applied, which provides a wide range of voltages with low harmonic distortion.

The paper is organized in the following sequence, section II explains the SHE-PWM method in the base case and the evolutionary algorithms are introduced to solve the nonlinear equations. In section III, the SHE technique in the proposed method is introduced to overcome the problems are arising in section II, and accordingly, the simulation results are reported in section IV.

## II. SHE-PWM METHOD IN BASE CASE

The typical structure of a single-phase H-bridge inverter is shown in Fig. 1, which consists of a series connecting single-phase H-bridge cells.

In most studies, the voltage waveform in the MLIs is studied by the SHE-PWM method. The inverter output voltage waveform has the symmetry of a quarter of an individual wave. By considering the DC-link source voltage as common, the Fourier series extension of the output voltage waveform is written as follows:

$$V(\omega t) = \sum_{n=1}^{\infty} V_n \sin(n\omega t) \quad (1)$$

Due to odd quarter-wave symmetry, harmonics with even order become zero. The switching angles are limited between zero and $\frac{\pi}{2}$. As a result, $V_n$ can be expressed as follows:

$$V_n = \frac{4V_{dc}}{\pi} \sum_{n=1,3,5,...}^{\infty} \frac{1}{n} \begin{Bmatrix} \cos(n\theta_1) - \cos(n\theta_2) + \\ \cos(n\theta_3) + \cos(n\theta_4) - \\ \cos(n\theta_5) - \cos(n\theta_6) \end{Bmatrix} \sin(n\omega t) \quad (2)$$

According to 3, six switching angles will be calculated. The first one is solved for the fundamental component, and others are solved to eliminating the lower order harmonic component. The 3rd, 5th, 7th, 9th, and 11th harmonics have been selected for elimination in the studied system.

$$V_1 = \frac{4V_{dc}}{\pi} \begin{Bmatrix} \cos(n\theta_1) - \cos(n\theta_2) + \cos(n\theta_3) + \\ \cos(n\theta_4) - \cos(n\theta_5) - \cos(n\theta_6) \end{Bmatrix}$$

$$V_3 = \frac{4V_{dc}}{3\pi} \begin{Bmatrix} \cos(3\theta_1) - \cos(3\theta_2) + \cos(3\theta_3) + \\ \cos(3\theta_4) - \cos(3\theta_5) - \cos(3\theta_6) \end{Bmatrix}$$

.

.

$$V_{11} = \frac{4V_{dc}}{11\pi} \begin{Bmatrix} \cos(11\theta_1) - \cos(11\theta_2) + \cos(11\theta_3) + \\ \cos(11\theta_4) - \cos(11\theta_5) - \cos(11\theta_6) \end{Bmatrix}$$

Also, with component $V_1$, the modulation index ($M$) can be introduced as follows:

$$M = \pi \times \frac{V_1}{2V_{dc}} \qquad (0 \le M \le 1) \quad (4)$$

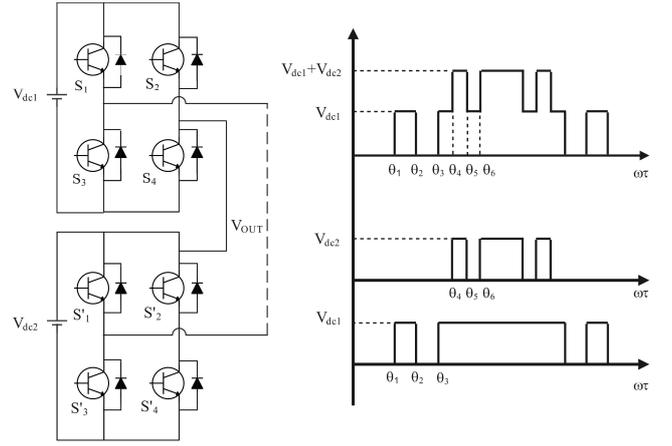

Fig. 1. A single-phase five-level CHB inverter. (a) Typical structure of a single-phase five-level CHB inverter. (b) Typical five-level SHE-PWM waveform.

According to (4), depending on the value of $V_1$, $M$ varies between 0 and 1. To executing the SHE-PWM principle, achieving the desired AC output voltage, and eliminating low-order harmonics, a cost function is defined as follows:

$$f = A \times \left| M - \frac{|V_1|}{sV_{dc}} \right| + B \times \sum_{i=2}^{ks} \frac{1}{h_{si}} \frac{|V_{si}|}{sV_{dc}} \quad (5)$$

$$i = 2,3,...,s$$

By solving (5), for each given value of the modulation index, the switching angles are obtained in a five-level inverter.

In this research, $s$ is used for the CHB inverter to indicate the separated DC sources and the number of voltages levels ($2s+1$). Hence, the number of harmonicas that can be eliminated from the output voltage is equal to ($s-1$) If the number of switches at each output voltage level is $k$, then the switching frequency of the SHE-PWM method will be increased to $ks$.

Several methods have been proposed to solve the nonlinear transcendental equations of the SHE-PWM method. The main objective of SHE-PWM is to calculate the appropriate switching angles so that the selected harmonic components are eliminated and at the same time the amplitude of the fundamental output voltage reaches the desired value. The proposed new methods are the use of evolutionary algorithms. One of the advantages of these methods is the ability to calculate for a wide range of modulation indices.

These methods can be used for equations with any number of inverter surfaces. Because it is simpler and more practical, the PSO algorithm is proposed in this paper.

## III. SHE TECHNUQE IN PROPOSD METHOD

The SHE method has a significant weakness in that the solution does not exist in all ranges of the modulation index. This means that in some modulation indices, there are no optimal switching angles. To overcome this disability, a proposed method is introduced to vary the DC-link voltage.

Hence, a high-frequency isolated DC-DC converter is used to perform such an approach. Therefore, in some cases where the

SHE equations do not have a response for a certain range of the modulation index, especially at low modulation index, the proposed method can be an effective solution to the problem. The use of high-frequency isolated DC-DC converters changes the voltage range of DC-links of the CHB cells. In the proposed method, the DC-DC converter is controlled in such a way that the modulation index can be increased equal to 1. The proposed scheme for increasing the modulation index using the SHE-PWM method is shown in Fig. 2.

The following relationships formulate the proposed idea. The per-unit value of the output voltage in this proposed method can be defined as (6).

$$V_{o,pu} = \frac{V_1}{V_{dc1} + V_{dc2}} \qquad (6)$$

$$(V_{dc1} = V_{dc2} \Rightarrow 2V_{dc} = V_{DC}) \qquad (7)$$

$$V_{o,pu} = \frac{V_1}{V_{DC}} \qquad (8)$$

Equation (8) can be obtained using (7) and (6). The relationships between the modulation index in the conventional method and the proposed method can be defined as:

$$\begin{cases} M_{old} = \dfrac{V_1}{V_{DC}} \\ M_{new} = \dfrac{V_1}{DV_{DC}} \end{cases} \Rightarrow \begin{cases} V_1 = M_{old} \times V_{DC} \\ V_1 = M_{new} \times DV_{DC} \end{cases} \qquad (9)$$

and thereby:

$$M_{old} = D \times M_{new} \xrightarrow{M_{new}=1} M_{old} = D \qquad (10)$$

Equation (10) shows the relationship between $D$ (Full Bridge Converter Duty Cycle) and the conventional mode modulation index. By changing $D$ between 0 and 1, the desired per unit value of voltage can be obtained. The new switching angles are collected in a lookup table obtained from the SHE technique.

## IV. SIMULATION RESULTS

To prove the proposed method and verify the theoretical results, the studied system is simulated in MATLAB SIMULINK. Initially, the SHE technique formulas are applied to the PSO algorithm, then the angles are calculated. In the SIMULINK MATLAB, a 5-level CHB inverter is simulated, then the angles obtained from the PSO algorithm are applied in the Simulink MATLAB simulated inverter software. The nominal voltage for each DC source is considered 200 volts. The simulation results for the modulation indexes are presented in Table I.

According to Fig. 3 and Fig. 4, by analyzing the output voltage waveform Fast Fourier transform (FFT), it can be seen that in the conventional method, the obtained voltage has low-order harmonics, and in other words, it has a high THD amount. However, in the proposed method, the low-order harmonics are eliminated or greatly attenuated.

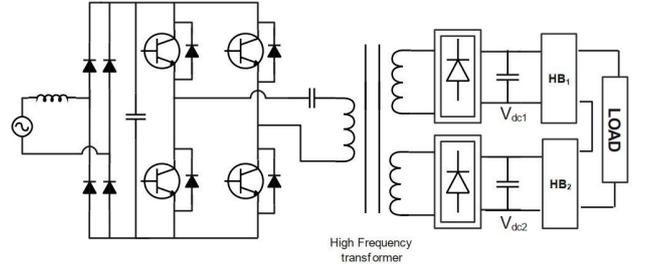

Fig. 2. Suggested a schematic of CHB inverter fed by high frequency isolated DC-DC converter.

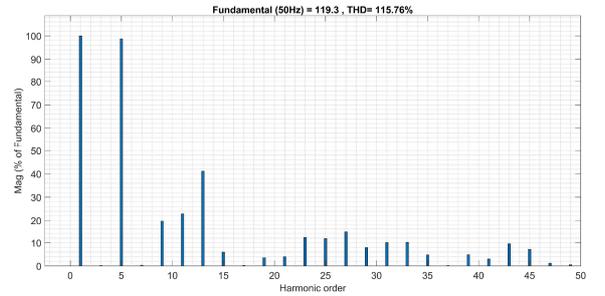

Fig. 3. FFT analysis of inverter phase voltage waveform in $V_{o,pu}$=0.3 in the conventional method.

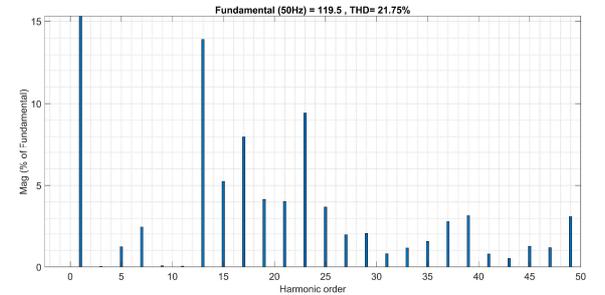

Fig. 4. FFT analysis of inverter phase voltage waveform in $V_{o,pu}$=0.3 in the proposed method.

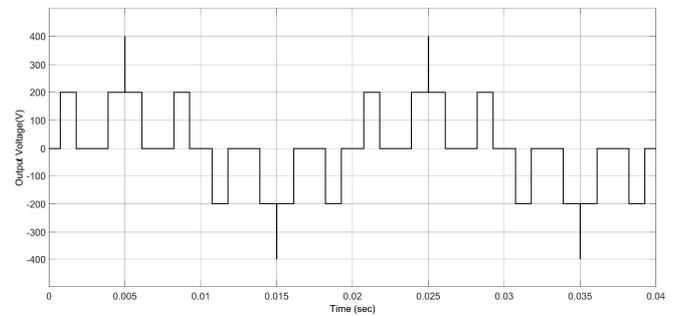

Fig. 5. Inverter phase output voltage waveform $V_{o,pu}$=0.3 in the conventional method.

According to Fig. 7 and Fig. 8, the amount of selected low-order harmonics in the SHE method is properly eliminated by applying the proposed idea and the quality of the obtained.

In the proposed method, not only the problem of low-order harmonic levels in the low-voltage unit is solved, but even in the high-voltage unit, the improvement in voltage quality is significant. Using this new method, we achieve a wide voltage range in which the selected unique harmonics are eliminated or attenuated.

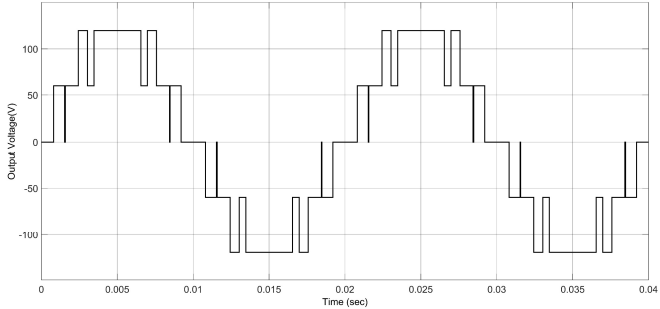

Fig. 6. Inverter phase output voltage waveform in $V_{o,pu}$ =0.3 in the proposed method.

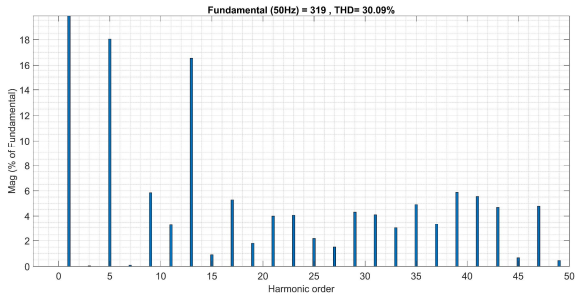

Fig. 7. FFT analysis of inverter phase voltage waveform in $V_{o,pu}$ =0.8 in the conventional method.

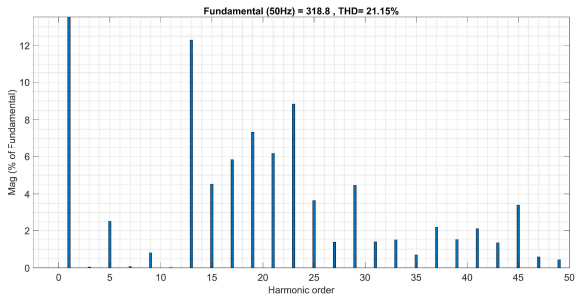

Fig. 8 FFT analysis of inverter phase voltage waveform $V_{o,pu}$ =0.8 in the proposed method.

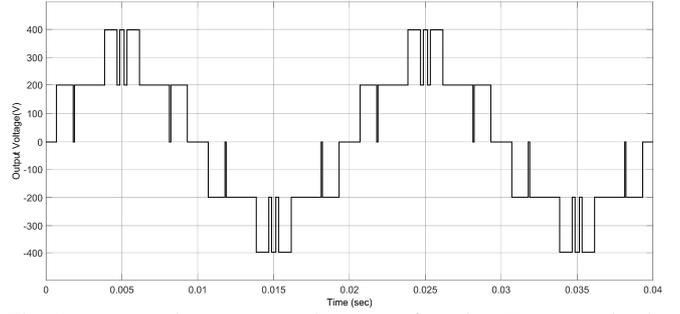

Fig. 9. Inverter phase output voltage waveform in $V_{o,pu}$ = 0.8 in the conventional method.

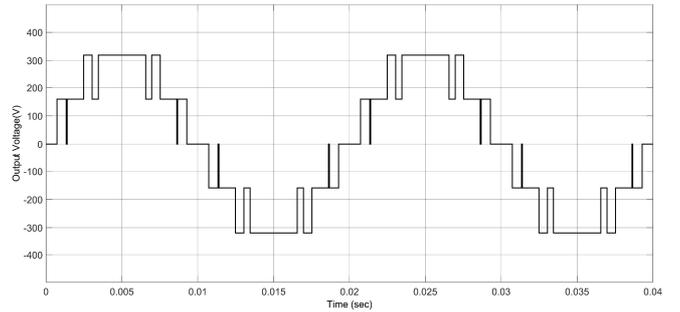

Fig. 10. Inverter phase output voltage waveform $V_{o,pu}$ =0.8 in the proposed method.

Table I. Comparison between the results obtained for the conventional method and the proposed method.

| no | per unit value of voltage | Voltage THD% conventional | Voltage THD% proposed the idea | The rate of improvement |
|---|---|---|---|---|
| 1 | 0.1 | 216.57 | 23.46 | 89.16% |
| 2 | 0.2 | 141.58 | 25.58 | 81.93% |
| 3 | 0.3 | 115.76 | 21.75 | 81.21% |
| 4 | 0.4 | 71.23 | 22.22 | 68.8% |
| 5 | 0.5 | 44.81 | 20.75 | 53.69% |
| 6 | 0.6 | 32.07 | 21.75 | 32.17% |
| 7 | 0.7 | 35.73 | 22.25 | 37.72% |
| 8 | 0.8 | 30.09 | 21.15 | 28.84% |
| 9 | 0.9 | 33.5 | 23.62 | 29.49% |
| 10 | 1 | 20.75 | 20.75 | - |

## V. Conclusion

In this paper, a method based on the use of the SHE-PWM technique for a CHB MLI to increase the output voltage's quality is introduced. This method is based on varying the DC voltage in the output voltage function. Simulation studies confirmed the correct operation of the proposed method. As the results showed, the harmonic distortion of the proposed method's output voltage is significantly reduced. For example, by examining the simulation results, the THD output voltage index for both methods investigated, including the conventional method and the proposed method at 0.2 $V_{o,pu}$ were 141.58% and 25.58%, respectively significant improvement in the quality of the voltage obtained. In other words, the proposed method for harmonic distortion in all voltage ranges, in particular, greatly improves on low voltage ranges.

# VI. References

[1] E. Babaei, Y. Azimpour and M. Farhadi Kangarlu, "Charge balance control of a seven-level asymmetric cascade multilevel inverter," 2012 3rd IEEE International Symposium on Power Electronics for Distributed Generation Systems (PEDG), Aalborg, 2012, pp. 674-681, doi:10.1109/PEDG.2012.6254075.

[2] M. Vijeh, M. Rezanejad, E. Samadaei, and K. Bertilsson, "A general review of multilevel inverters based on main submodules: Structural point of view," IEEE Transactions on Power Electronics, vol. 34, no. 10, pp. 9479–9502, Oct. 2019.

[3] L. Tapia Hector Jua, A. Rodriguez Jose Juan, D. Gonzalez Aurelio, and R. Resendiz Juvenal, "Eight levels multilevel voltage source inverter modulation technique," IEEE Latin America Transactions, vol. 16, no. 4, pp. 1121–1127, Apr. 2018.

[4] S. Maurya, D. Mishra, K. Singh, A. Mishra, and Y. Pandey, "An efficient technique to reduce total harmonics distortion in cascaded hbridge multilevel inverter," in 2019 IEEE International Conference on Electrical, Computer and Communication Technologies (ICECCT). IEEE, Oct. 2019, pp. 1–5.

[5] ZAMBRA D A B, RECH C, PINHEIRO J R. "Comparison of neutral point-clamped, symmetrical, and hybrid asymmetrical multilevel inverters. "IEEE Transactions on Industrial Electronics, 2010, 57(7): 2297-2306

[6] A. Perez-Basante, S. Ceballos, G. Konstantinou, J. Pou, I. Kortabarria, and I. M. d. Alegr'ıa, "A universal formulation for multilevel selectiveharmonic-eliminated pwm with half-wave symmetry," IEEE Transactions on Power Electronics, vol. 34, no. 1, pp. 943–957, Jan. 2019.

[7] K. K. Gupta, A. Ranjan, P. Bhatnagar, L. K. Sahu and S. Jain, "Multilevel Inverter Topologies with Reduced Device Count: A Review," in IEEE Transactions on Power Electronics, vol. 31, no. 1, pp. 135-151, Jan. 2016, doi: 10.1109/TPEL.2015.2405012.

[8] M. Farhadi Kangarlu and E. Babaei, "A Generalized Cascaded Multilevel Inverter Using Series Connection of Submultilevel Inverters," in IEEE Transactions on Power Electronics, vol. 28, no. 2, pp. 625-636, Feb. 2013, doi: 10.1109/TPEL.2012.2203339.

[9] M. Sadoughi, A. Zakerian, A. Pourdadashnia and M. Farhadi-Kangarlu, "Selective Harmonic Elimination PWM for Cascaded H-bridge Multilevel Inverter with Wide Output Voltage Range Using PSO Algorithm," 2021 IEEE Texas Power and Energy Conference (TPEC), College Station, TX, USA, 2021, pp. 1-6, doi: 10.1109/TPEC51183.2021.9384945.

[10] A. Pourdadashnia, M. Farhadi-Kangarlu, and M. Sadoughi, "Staircase selective harmonic elimination in multilevel inverters to achieve wide output voltage range," 2021.

[11] Y. Zhang, Z. Zhao and J. Zhu, "A Hybrid PWM Applied to High-Power Three-Level Inverter-Fed Induction-Motor Drives," in IEEE Transactions on Industrial Electronics, vol. 58, no. 8, pp. 3409-3420, Aug. 2011, doi: 10.1109/TIE.2010.2090836.

[12] M. Wu, K. Wang, K. Yang, G. Konstantinou, Y. W. Li and Y. Li, "Unified Selective Harmonic Elimination Control for Four-Level Hybrid-Clamped Inverters," in IEEE Transactions on Power Electronics, vol. 35, no. 11, pp. 11488-11501, Nov. 2020, doi: 10.1109/TPEL.2020.2985090.

[13] A. Pérez-Basante, S. Ceballos, G. Konstantinou, J. Pou, I. Kortabarria and I. M. d. Alegría, "A Universal Formulation for Multilevel Selective-Harmonic-Eliminated PWM With Half-Wave Symmetry," in IEEE Transactions on Power Electronics, vol. 34, no. 1, pp. 943-957, Jan. 2019. doi: 10.1109/TPEL.2018.2819724.

[14] J. Kennedy, R.C. Eberhart, "Particle Swarm Optimization," Proc. IEEE Int. of. Neural Networks, Piscataway, NJ, USA, 1942-1948, 1995.

[15] H. Haggi, W. Sun and J. Qi, "Multi-Objective PMU Allocation for Resilient Power System Monitoring," 2020 IEEE Power & Energy Society General Meeting (PESGM), 2020, pp. 1-5, doi: 10.1109/PESGM41954.2020.9281963.

[16] A. Taghavirashidizadeh, R. Parsibenehkohal, M. Hayerikhiyavi, and M. Zahedi, "A genetic algorithm for multi-objective reconfiguration of balanced and unbalanced distribution systems in fuzzy framework," Journal of Critical Reviews, vol. 7, no. 7, pp. 639–343, 2020.

[17] M. Etesami, N. Ghasemi, D. M. Vilathgamuwa, W. L Malan, "Particle swarm optimisation-based modified SHE method for cascaded Hbridge multilevel inverters", IET Power Electronics, Vol. 10, No. 1, pp. 18-28, 2017

[18] H. Taghizadeh and M. Tarafdar Hagh, "Harmonic Elimination of Cascade Multilevel Inverters with Nonequal DC Sources Using Particle Swarm Optimization," in IEEE Transactions on Industrial Electronics, vol. 57, no. 11, pp. 3678-3684, Nov. 2010, doi: 10.1109/TIE.2010.2041736.

[19] H. Taghizadeh and M. Tarafdar Hagh, "Harmonic elimination of multilevel inverters using particle swarm optimization," 2008 IEEE International Symposium on Industrial Electronics, Cambridge, 2008, pp. 393-396, doi: 10.1109/ISIE.2008.4677093.

[20] M. T. Hagh, H. Taghizadeh, and K. Razi, "Harmonic minimization in multilevel inverters using modified species-based particle swarm optimization," IEEE Transactions on Power Electronics, vol. 24, no. 10, pp. 2259–2267, Oct. 2009.

[21] M. Farhadi Kangarlu and E. Babaei, "Variable DC voltage as a solution to improve output voltage quality in multilevel converters," 4th Annual International Power Electronics, Drive Systems and Technologies Conference, Tehran, 2013, pp. 242-247, doi: 10.1109/PEDSTC.2013.6506711.